\documentclass[aps,prb,superscriptaddress,footinbib,showpacs,twocolumn]{revtex4}

\usepackage[ansinew]{inputenc}

\usepackage{xspace}
\usepackage[dvips]{graphicx}
\newcommand{\invcm}{cm\ensuremath{^{-1}}\xspace}
\newcommand{\cBN}{\hbox{\textit{c}-BN}\xspace}
\newcommand{\Sam}{\ensuremath{\mathrm{SrB}_4\mathrm{O}_7\mathrm{:Sm}^{2\mathrm{+}}}}
\newcommand{\degree}{\ensuremath{^\circ}\xspace}
\newcommand{\micron}{\ensuremath{\mu}m\xspace}

\newcommand{\ten}[1]{\ensuremath{10^{#1}}}
\newcommand{\Vunit}{\AA$^3$/at\xspace}
\newcommand{\etal}{\emph{et al.}\xspace}
\newcommand{\gammath}[1]{\ensuremath{\gamma_{th#1}}\xspace}
\newcommand{\PVT}{$P$--$V$--$T$\xspace}
\newcommand{\OLC}[1]{Ref.~\onlinecite{#1}\xspace}
\newcommand{\nuTO}{\ensuremath{\nu^{TO}}}
\newcommand{\gammaTO}{\ensuremath{\gamma^{TO}}}
\begin{document}

\title{Equation of state of cubic boron nitride at high pressures and temperatures}

\author{F. Datchi}
\email[E-mail: ]{datchi@impmc.jussieu.fr} \affiliation{Université Pierre et Marie Curie-Paris 6, UMR 7590, 140
rue Lourmel, Paris, F-75015, France; CNRS, UMR 7590, Paris, F-75015, France.}
\author{A. Dewaele}
\affiliation{DIF/Département de Physique Théorique et Appliquée, CEA, BP12, 91680 Bruyères-Le-Châtel, France}
\author{Y. Le Godec}
\affiliation{Université Pierre et Marie Curie-Paris 6, UMR 7590, 140 rue Lourmel, Paris, F-75015, France; CNRS,
UMR 7590, Paris, F-75015, France.}
\author{P. Loubeyre}
\affiliation{DIF/Département de Physique Théorique et Appliquée, CEA, BP12, 91680 Bruyères-Le-Châtel, France}
\date{\today}

\begin{abstract}
We report accurate measurements of the equation of state (EOS) of cubic boron nitride by x-ray diffraction up to
160 GPa at 295 K and 80 GPa in the range 500-900 K. Experiments were performed on single-crystals embedded in a
quasi-hydrostatic pressure medium (helium or neon). Comparison between the present EOS data at 295~K and
literature allows us to critically review the recent calibrations of the ruby standard. The full $P$-$V$-$T$ data
set can be represented by a Mie-Gr\"{u}neisen model, which enables us to extract all relevant thermodynamic
parameters: bulk modulus and its first pressure-derivative, thermal expansion coefficient, thermal Gr\"{u}neisen
parameter and its volume dependence. This equation of state is used to determine the isothermal Gr\"{u}neisen
mode parameter of the Raman TO band. A new formulation of the pressure scale based on this Raman mode, using
physically-constrained parameters, is deduced.

\end{abstract}


\maketitle
\section{Introduction}

Cubic boron nitride (\cBN) is a material with remarkable properties: extreme hardness, chemical inertness, large
band gap and high mechanical and thermal stabilities. This makes it very attractive for a number of applications
including abrasive or protective-coating material and microelectronic devices. Knowing the properties of \cBN
under conditions of high pressure ($P$) and temperature ($T$) is important for some of these applications. It has
also been recognized\cite{Aleksandrov89, Eremets96} that c-BN could be a useful pressure sensor in high pressure
(HP)- high temperature (HT) experiments in diamond anvil cells, based on its intense Raman TO band (1054 \invcm
at ambient conditions) which is well separated from the signal of the anvils. This has recently motivated
investigations of the Raman spectra under extreme conditions of $P$ and $T$\cite{Datchi2004, Kawamoto2004,
Goncharov2005}.

As pointed out by Holzapfel\cite{Holzapfel2003}, the accurate determination of the equation of state (EOS) of
solids with very large bulk modulus is a stringent test for the ruby sensor calibration, a matter which has been
under intensive debate recently\cite{Occelli2003, Holzapfel2003, Kunc2003, Dewaele2004, Chijioke2005,
Holzapfel2005, Dorogokupets2007}. For highly incompressible solids like \cBN, experiments in the megabar range
are required in order to extract reliable values of the bulk modulus $B_0$ and its first pressure derivative
$B'_0$. At ambient temperature, the equation of state has been so far investigated up to 34 GPa in a He pressure
medium\cite{Aleksandrov89}, to 66 GPa in N$_2$\cite{Solozhenko1998}, and 120 GPa in
methanol-ethanol\cite{Knittle89}. Although the maximum pressure in the latter study is relatively high, the small
number of measurements, their limited accuracy, and the use of a pressure media prone to nonhydrostatic stress
above $\sim 20$ GPa\cite{Takemura2001} reflects in poorly constrained and questionable values of $B_0$ and
$B'_0$.

There is to date no available EOS data at simultaneous high pressure and temperature. By contrast, several theoretical
studies\cite{Albe97,Kern1999,Tohei2006} have reported thermal properties at ambient and high pressures. These works
relied on the assumption that \cBN can be treated as a quasi-harmonic solid; this still needs to be validated by
experiment.

We present here careful and accurate measurements of the EOS up to 160 GPa at 295 K and up to 80 GPa at high
temperatures (500-900 K). We used helium or neon (at high T) as pressure transmitting medium since they are known
to provide the closest approximation to hydrostatic compression in the megabar range\cite{Takemura2001,
Dewaele2004}. The thermal expansion at ambient pressure has also been investigated to 950 K. We show that our
data can be represented by a simple Mie-Gr\"{u}neisen model. This allows us to extract all the relevant
thermodynamic parameters. The EOS is then used to determine the isothermal Gr\"{u}neisen parameter of the Raman
TO band. Accordingly, the pressure-scale based on this Raman band is reformulated using physical constraints.

This article is organized as follows: in section II, we present the experimental procedure; section III is
devoted to the presentation of the room-temperature equation of state; the thermal expansion at room pressure
follows in section IV; the presentation and analysis of \PVT data at simultaneous HP-HT are given in section V;
in section VI, the thermal dependence of the isothermal Gr\"{u}neisen parameter of the Raman TO band is
investigated and a new formulation of the Raman pressure scale is then presented. Section VII finally gives
concluding remarks.

\section{Experimental procedure}

The present experiments were conducted with single crystals of \cBN of size ranging from 3 to 15 \micron. These
were selected from a powder batch using their Raman signal as a criterium for good crystallinity: crystals which
presented intense Raman first-order TO and LO bands with lorentzian shapes and small bandwidths ($\sim5$~\invcm)
were selected. A few crystals were then loaded into the experimental volume of a diamond anvil cell, along with
ruby and \Sam as pressure sensors. Care was taken in order to position the samples and the pressure sensors
within a few micron distance to each other and at the center of the diamond culet.
   We used diamond anvils with flat culets of 0.1 to
0.4 mm and rhenium gaskets. Helium was chosen as pressure transmitting medium for the room temperature
experiments above 50 GPa and neon was used otherwise. In the $P-T$ range of the present experiments, these two
pressure transmitting media are known to provide the best approximation to hydrostatic
conditions\cite{Takemura2001, Dewaele2004}.

 Membrane diamond anvil cells (MDAC) designed for high-temperature operation were used. The
cells could be fitted as a whole inside a ring-shaped resistive heater. To achieve temperatures above 800 K, a
smaller, additional heater made of a resistive wire coiled around a ceramic tube, is positioned around the
anvil-gasket assembly. The temperature of the heaters are regulated within 1 K using commercial devices. An
isolated, K-type thermocouple is fixed by ceramic cement with its head in contact with the diamond anvil, close
to the gasket. The ensemble is heated in air or in a Ar-H$_2$(2\%) reducing atmosphere. Numerous previous
experiments have shown that the temperature measured by the thermocouple is within 5 K of the sample temperature
\cite{Datchi2000,Datchi2004,Giordano2006}.

Pressure was determined using the pressure shift of the luminescence lines of ruby (at 295~K) or \Sam. The
pressures reported here are based on Holzapfel's 2005 ruby scale\cite{Holzapfel2005}, hereafter denoted H2005. We
also compared the results obtained with other ruby
calibrations\cite{Mao86,Holzapfel2003,Chijioke2005,Holzapfel2005,Dorogokupets2007} and, as discussed below, H2005
was found to provide excellent consistency between present and literature data. The calibration of the \Sam
sensor\cite{Datchi97}, initialy based on the ruby scale from Ref.~\onlinecite{Mao86},  was also modified to match
the H2005 scale. For the measurements above 100~GPa at 295~K, the ruby signal was too weak, thus the pressure was
determined from the equation of state of $^4$He (Ref. \onlinecite{Loubeyre1998}, with proper correction for the
H2005 ruby scale). The volume of $^4$He was calculated using the reflections present on the same diffraction
patterns as the \cBN sample and excellent agreement between this pressure determination and that from ruby was
observed below 100~GPa.

Angular-dispersive x-ray diffraction experiments were performed on beamline ID27 of the European Synchrotron
Radiation Facility (ESRF, Grenoble, France). The monochromatic beam ($\lambda=0.3738$~\AA) was focused to a
$\approx 7\times10$~$\mathrm{\mu m^2}$ spot. Diffracted x-rays were collected by a MAR345 image plate while the
MDAC was continuously rotated about the $\phi$-axis by $\pm$20\degree. The images were integrated using the fit2D
program\cite{fit2D}. Between 6 and 9 single-crystal reflections could be observed depending on the sample, with a
resolution up to $\approx0.83$~\AA. All of them could be indexed in the zinc-blende structure (space group
$F\overline{4}3m$) reported for this material\cite{Wentorf1957}. The lattice parameter $a$ was refined with the
program UnitCell\cite{UnitCell}, using the measured $d$-spacings of all observed reflections. The uncertainty on
$a$ was on average $5\times10^{-4}$~\AA.

\section{The room-temperature equation of state}

\begin{figure}[tb]
\includegraphics[width=8.5cm]{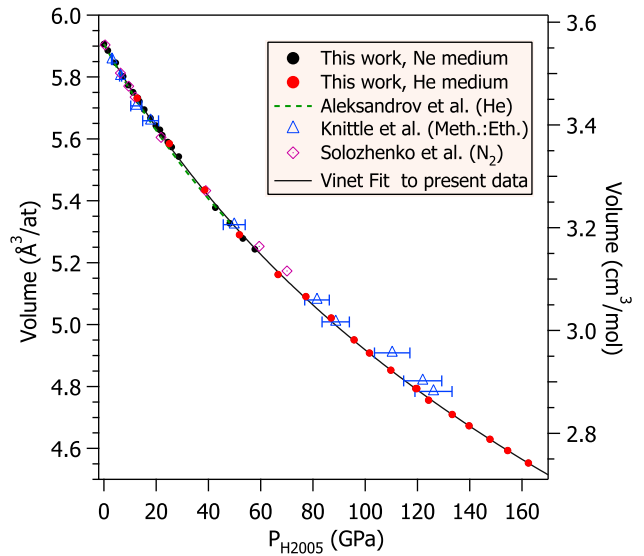}
\caption{\label{RTEOS}(Color online) Volume per atom of \cBN as a function of pressure  at 295 K. The open and
solid circles are present measurements on samples in a neon and helium pressure medium respectively. The
estimated error bars are within the symbol sizes. Triangles: \OLC{Knittle89} (in methanol/ethanol); dashed line:
\OLC{Aleksandrov89} (in helium); squares: \OLC{Solozhenko1998} (in nitrogen). The solid line is the fit to the
present data using the Vinet equation \cite{Vinet1987}[$V_0=5.9062(6)$ \AA$^3$/at, $B_0=395(2)$~GPa,
$B'_0=3.62(5)$].}
\end{figure}

\begin{figure}[tb]
\includegraphics[width=6.5cm]{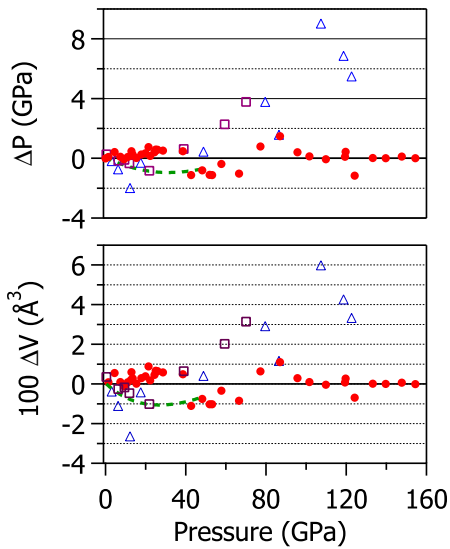}
\caption{\label{DVDP} (Color online) Comparison between experimental data at 295 K from this work (circles),
Ref.~\onlinecite{Knittle89} (triangles), \OLC{Solozhenko1998} (squares) and Ref.~\onlinecite{Aleksandrov89}
(dashed line). The fit to the present data using the Vinet EOS\cite{Vinet1987} is used as the reference. The top
graph shows the difference in pressure and the bottom one, the difference in volume multiplied by a factor 100.}
\end{figure}

\begin{table}
\caption{\label{RTEOSfits} Comparison between values for the zero-pressure isothermal bulk modulus $B_0$ and its
first pressure derivative $B'_0$ obtained by fitting the present data at ambient temperature to different EOS
models: Vinet\cite{Vinet1987}, second-order Birch-Murnaghan (BM)\cite{Birch1978} and Holzapfel's AP2
form\cite{Holzapfel2001}. The zero-pressure volume $V_0$(295 K) was fixed to 5.9062~\Vunit.}
\begin{ruledtabular}
\begin{tabular}{cccc}
  Model         & Vinet     & BM      & AP2      \\
 $B_0$, GPa     & 395(2)    & 396(2)  & 397(2)\\
 $B'_0$         & 3.62(5)   & 3.54(5) & 3.50(5)\\
 \end{tabular}
 \end{ruledtabular}
 \end{table}

The volume per atom $V$ of \cBN  at 295~K ($V=a^3/8$) was measured up to 57.8 GPa in a neon pressure medium and
up to 162 GPa in a helium pressure medium. Excellent agreement is observed between the two data sets. The results
are gathered in Fig.~\ref{RTEOS} and Table~\ref{PVTdata}. As expected, no phase transition was observed in this
pressure range.

To extract the values of the zero-pressure isothermal bulk modulus $B_0$ and its first-pressure derivative $B'_0$
from the present data, we considered three different phenomenological models of equation of state: the Vinet
form\cite{Vinet1987}, the Birch-Murnaghan second-order form\cite{Birch1978} and Holzapfel's AP2 equation
\cite{Holzapfel2001}. These relations were least-squares fitted to the data. The results are listed in
Table~\ref{RTEOSfits} and the fit using the Vinet equation\cite{Vinet1987} is plotted in Fig.~\ref{RTEOS}. In the
compression range probed here ($0.77\leq V/V_0\leq 1$) the three models fit equally well the data and give very
similar values for $B_0$ ($395(2)\div397(2)$~GPa) and its first pressure-derivative $B'_{0}$
($3.50(5)\div3.62(5)$). Here as in all the text, the number in parenthesis indicate the standard deviation for
the last digit of the fitting parameter; it does not reflect its absolute uncertainty, which primarily depends
here on the one of the ruby standard. The volume and pressure difference between experimental data and the Vinet
fit is shown in Fig.~\ref{DVDP}. The volume $V_0$ at $P=0$ and 295 K was taken as the measured value at 1 atm of
$5.9062(6)$~\AA$^3$/at, corresponding to a lattice parameter $a_0$ of $3.6152(2)$~\AA. The latter values are
identical, within uncertainties, to previous reports \cite{Slack75,Aleksandrov89,Knittle89}.

Figs.~\ref{RTEOS} and~\ref{DVDP} also shows experimental data obtained in previous
studies\cite{Aleksandrov89,Knittle89}. For comparison with the present work, the original pressure values have
been rescaled onto the H2005 scale. Aleksandrov et al.\cite{Aleksandrov89} reported their measurements up to 34
GPa in He in the form of a quadratic fit of the compression factor $\Delta\rho/\rho_0$--where $\rho$ is the
density, vs. the relative shift of the ruby $R_1$ line. Their absolute volume values differ from ours by less
than $2\times\ten{-2}$~\Vunit, corresponding to a maximum pressure difference of 2~GPa. By contrast, systematic
deviations are observed between our measurements and those of Solozhenko \etal\cite{Solozhenko1998} and Knittle
\emph{et al.}\cite{Knittle89} above about 40 GPa, where their measured volume at a given pressure is
systematically higher than our. This is likely due, at least in part, to the larger nonhydrostatic stresses
produced by the N$_2$ or methanol-ethanol pressure medium used in Solozhenko \etal\cite{Solozhenko1998} and
Knittle \etal\cite{Knittle89} experiments respecitvely.

As mentioned above, the values of $B_0$ and $B'_0$ obtained by fitting the EOS of highly incompressible solids
like \cBN and diamond are very sensitive to the chosen pressure calibration for the ruby standard. As a matter of
fact, the first hint that the commonly used Mao \etal 's 1986 calibration\cite{Mao86} (hereafter denoted as
MXB1986) becomes increasingly wrong with pressure was given by the measurements of the diamond EOS by Aleksandrov
\etal\cite{Aleksandrov89} to 40~GPa in 1989. More recently, the very accurate measurements of the diamond EOS up
to megabar pressures by Occelli \etal\cite{Occelli2003}  evidenced more firmly the discrepancy between the value
found for $B'_0$ using ultrasonic techniques ($4.0\pm0.7$), on one hand, and that  obtained from the EOS (3.0)
when pressure is calculated with the MXB1986 scale, on the other hand. Consequently, a revision of the latter
scale has been proposed by several authors\cite{Holzapfel2003, Kunc2003,Dewaele2004, Chijioke2005, Holzapfel2005,
Dorogokupets2007}, based on the diamond data and/or those recently obtained for the EOS of several metals that
showed the same trend\cite{Dewaele2004}.

\begin{table}
 \caption{\label{RubyScales} Comparison between values of $B_0$ (in GPa) and $B'_0$ of \cBN and natural diamond (C) at 295~K obtained with
different ruby scales from the literature. $B_0$ and $B'_0$ are determined by fitting the present data (\cBN) and
Occelli \etal 's data (C) to the Vinet model\protect\cite{Vinet1987}. In all fits, the values of the
zero-pressure volume was fixed to 5.9062~\Vunit (\cBN) and 5.6733~\Vunit (C). $\chi^2$ indicates the goodness of
fit.}
\begin{ruledtabular}
\begin{tabular}{ccccccc}
 & \multicolumn{3}{c}{\cBN} & \multicolumn{3}{c}{Diamond}\\
 Ruby scale & $B_0$ & $B'_0$ & $\chi^2$ & $B_0$ & $B'_0$ & $\chi^2$ \\
 \hline
MXB1986\cite{Mao86}                 &   397(3)  &   2.75(6) &   14.2    &   447(3)  &   3.00(7) &   4.5 \\
AGSY1989\cite{Aleksandrov89}        &   394(3)  &   3.93(7) &   21      &   440(3)  &   4.38(8) &   6.4 \\
H2003\cite{Holzapfel2003}           &   390(2)  &   3.49(5) &   12.1    &   435(2)  &   3.87(4) &   1.5 \\
DLM2004\cite{Dewaele2004}            &   395(3)  &   3.28(6) &   16.8    &   440(3)  &   3.66(7) &   4.6 \\
H2005\cite{Holzapfel2005}           &   395(2)  &   3.62(2) &   12.8    &   443(3)  &   3.97(5) &   1.5 \\
CNSS2005\cite{Chijioke2005}         &   387(3)  &   3.64(7) &   18.2    &   432(3)  &   4.05(7) &   5.3 \\
DO2007\cite{Dorogokupets2007}       &   398(2)  &   3.35(5) &   13.6    &   444(2)  &   3.72(5) &   2.6\\
 \end{tabular}
 \end{ruledtabular}
 \end{table}

 \begin{table*}
\caption{\label{literature} Literature data for the values of $V_0$, $B_0$ and $B'_0$ of \cBN and diamond
($^{12}$C). $V_0$ is in \Vunit and $B_0$ in GPa. The various abbreviations are: DFT: density functional theory,
LDA: Local density approximation, GGA: generalized gradient approximation, ZPE: zero-point energy, DG:
Debye-Grüneisen model, LD: lattice dynamics, VMC: variational quantum Monte-Carlo, DMC: diffusion quatum Monte-Carlo; "static" stands for calculations of the static lattice.}
\begin{ruledtabular}
\begin{tabular}{cccccccccc}
 \multicolumn{5}{c}{\cBN} & \multicolumn{5}{c}{$^{12}$C}\\
 Reference          & $V_0$     & $B_0$     & $B'_0$ & Method                & Reference             & $V_0$  & $B_0$  & $B'_0$  & Method \\
 \hline
 \OLC{Grimsditch94}  &          & 400(20)    &       & Brillouin            & \OLC{McSkimin1972}    &        & 442(5)  & 4.0(7)  & Ultrasound\\
 \OLC{Albe97}        & 5.797    & 395       &  3.65  & DFT+LDA (static)     & \OLC{Vogelgesang96}   &        & 444.8(8)&         & Brillouin\\
                    & 5.884     & 387      & 3.66   & {\scriptsize DFT+LDA+ZPE (0~K)}    & \OLC{Pavone1993}      & 5.497  & 473     & 3.5     & DFT+LDA(static)  \\
                    & 5.888     & 385      & 3.66   & {\scriptsize DFT+LDA+ZPE+DG(300~K)}    & \OLC{Kunc2003}        & 5.510  & 465     & 3.63(3) & DFT+LDA (static) \\
 \OLC{Tohei2006}     & 5.745     & 391      &        & {\scriptsize DFT+LDA+LD (300~K)}   &                       & 5.697  & 433(2)  & 3.67(3) & DFT+GGA (static)\\
 \OLC{Karch1997}     & 5.788     & 397      & 3.6    & DFT+LDA (static)     & \OLC{Maezono2007}     & 5.529  & 454      & 3.65   &  {\scriptsize DFT+LDA+ZPE+QHA(300~K)}\\
 \OLC{Kern1999}      & 5.742     & 398      &        & DFT+LDA (static)     &                       & 5.722  & 422      &3.72    & {\scriptsize DFT+GGA+ZPE+QHA (300~K})\\
 \OLC{Furthmuller1994}& 5.718   & 397       & 3.59  &  DFT+LDA (static)     &                       & 5.604  & 472(4)   & 3.8(1) & {\scriptsize VMC+ZPE+QHA (300~K)}\\
 \OLC {Knittle89}    & 5.954    & 368       & 3.6   &  DFT+LDA (static)     &                       & 5.711  & 437(3)   & 3.7(1) & {\scriptsize DMC+ZPE+QHA (300~K)}\\
 \OLC{Xu1991}       & 5.905     & 370       & 3.8   & Tight-binding         & \\

 \end{tabular}
 \end{ruledtabular}
 \end{table*}

In Table~\ref{RubyScales} we compare the $B_0$ and $B'_0$ parameters obtained by fitting the Vinet EOS to the
present \cBN data using the various ruby calibrations proposed in the literature. For comparison and consistency
check, we also show the results obtained with the diamond data of Occelli \etal\cite{Occelli2003}. To decide
which ruby scale(s) seem(s) more reasonable, we list in Table \ref{literature} the values of $B_0$ and $B'_0$
determined by other means, whether experimental (ultrasonic or Brillouin scattering measurements) or theoretical.
Unfortunately, there exists only one determination of $B'_0$ for diamond using ultrasonic experiments under
pressure, which has a limited accuracy, and none has been reported yet for \cBN. The largest source of comparison
comes thus from the numerous theoretical studies based on first-principles ({\it ab initio}) techniques. Whereas
the bulk modulus and zero-pressure volume  show large variations among the various theoretical studies, the
values for $B'_0$ appear much less dependent on the theoretical approaches and approximations used in each of
them, especially if we consider the most recent works. Indeed, the two latest studies of diamond give values of
$B'_0$ between 3.65 and 3.8(1), whether the density functional theory (DFT), within LDA or GGA approximations, or
the quantum Monte-Carlo approach is used. Similarly, all DFT-LDA calculations on \cBN give $B'_0=3.6\pm0.05$,
irrespective of the considered pseudopotential. In the case of \cBN also, there is a nice agreement between the
latest studies on the values of $B_0$ which vary from 395 to 398 GPa, i.e in very good agreement with the
Brillouin scattering experiment\cite{Grimsditch94} ($400\pm20$ GPa). We see from Table \ref{RubyScales} that the
ruby scales H2003, H2005 and DO2007 are those giving the best consistency with available literature data, while
providing the best fits based on the least-squares $\chi^2$ criterium. The DLM2004 and CNSS2005 calibrations give
close results but a larger $\chi^2$, which could be ascribed to the used functional form. This also confirms that
the MXB1986 scale becomes increasingly wrong with pressure, underestimating it by about 8.8\% at 160 GPa. We note
that H2003, H2005, DLM2004, CNSS2005 and DO2007 scales agree within 3\% at 160 GPa, which may be considered as
the present uncertainty of the ruby scale at this pressure. It is now clear that independent experimental
determination of $B'_0$, such as given by sound-propagating experiments, for both \cBN and diamond could help to
better establish the ruby scale in the megabar range.

\begin{table}
 \caption{\label{PVTdata} Experimental $P-V-T$ data of \cBN obtained in the present work. $P$ is in GPa, $a$ in \AA, $V$ in
\AA$^3$/at and $T$ in K.}
\begin{ruledtabular}
\begin{tabular}{cccc@{\hspace{2mm}}|cccc}
    $P$     &  $T$  & $a$       & $V$           & $P$       &  $T$  & $a$       & $V$   \\\hline
    0.0     &   295 &   3.6152  &   5.907   &    0.0    &   377 &   3.6158  &   5.909   \\
    1.5     &   295 &   3.6110  &   5.886   &    0.0    &   479 &   3.6174  &   5.917   \\
    4.5     &   295 &   3.6029  &   5.846   &    9.8 &   497 &   3.5879  &   5.773   \\
    7.4     &   295 &   3.5938  &   5.802   &    21.6    &   499 &   3.5555  &   5.618   \\
    9.3     &   295 &   3.5880  &   5.774   &    2.0 &   500 &   3.6110  &   5.886   \\
    11.1    &   295 &   3.5832  &   5.751   &    25.1    &   500 &   3.5465  &   5.576   \\
    12.7    &   295 &   3.5792  &   5.732   &    13.9    &   500 &   3.5760  &   5.716   \\
    13.0    &   295 &   3.5793  &   5.732   &    46.5    &   501 &   3.4983  &   5.352   \\
    13.5    &   295 &   3.5771  &   5.721   &    44.5    &   501 &   3.5024  &   5.370   \\
    15.4    &   295 &   3.5715  &   5.695   &    0.0    &   573 &   3.6181  &   5.920   \\
    17.8    &   295 &   3.5658  &   5.667   &    39.3    &   600 &   3.5145  &   5.426   \\
    19.9    &   295 &   3.5607  &   5.643   &    47.9    &   600 &   3.4965  &   5.343   \\
    21.4    &   295 &   3.5578  &   5.629   &    56.5    &   600 &   3.4791  &   5.264   \\
    22.3    &   295 &   3.5541  &   5.612   &    66.3    &   600 &   3.4609  &   5.182   \\
    24.4    &   295 &   3.5494  &   5.590   &    76.2    &   600 &   3.4458  &   5.114   \\
    25.0    &   295 &   3.5484  &   5.585   &    84.2    &   600 &   3.4264  &   5.028   \\
    26.0    &   295 &   3.5461  &   5.574   &    0.0    &   674 &   3.6201  &   5.930   \\
    28.6    &   295 &   3.5395  &   5.543   &    8.2 &   748 &   3.5962  &   5.814   \\
    38.6    &   295 &   3.5164  &   5.435   &    4.6 &   748 &   3.6069  &   5.866   \\
    42.7    &   295 &   3.5040  &   5.378   &    18.9    &   749 &   3.5653  &   5.665   \\
    48.2    &   295 &   3.4930  &   5.327   &    35.8    &   749 &   3.5243  &   5.472   \\
    51.9    &   295 &   3.4849  &   5.290   &    42.1    &   750 &   3.5095  &   5.403   \\
    53.1    &   295 &   3.4823  &   5.278   &    49.5    &   750 &   3.4938  &   5.331   \\
    57.8    &   295 &   3.4746  &   5.244   &    14.7    &   750 &   3.5789  &   5.730   \\
    66.6    &   295 &   3.4565  &   5.162   &    0.0    &   775 &   3.6216  &   5.938   \\
    77.3    &   295 &   3.4405  &   5.091   &    0.0    &   873 &   3.6233  &   5.946   \\
    87.0    &   295 &   3.4249  &   5.022   &    54.2    &   900 &   3.4841  &   5.287   \\
    95.7    &   295 &   3.4087  &   4.951   &    0.0    &   948 &   3.6246  &   5.952   \\
    101.6   &   295 &   3.3989  &   4.908   \\
    109.8   &   295 &   3.3860  &   4.853   &\\
    119.3   &   295 &   3.3722  &   4.793   &\\
    119.7   &   295 &   3.3721  &   4.793   &\\
    124.3   &   295 &   3.3634  &   4.756   &\\
    133.3   &   295 &   3.3525  &   4.710   &\\
    139.7   &   295 &   3.3437  &   4.673   &\\
    147.7   &   295 &   3.3334  &   4.630   &\\
    154.5   &   295 &   3.3246  &   4.593   &\\
    162.5   &   295 &   3.3147  &   4.553   &\\

\end{tabular}
\end{ruledtabular}
\end{table}

\section{Thermal expansion at ambient pressure}

\begin{figure}[tb]
\includegraphics[width=8.5cm]{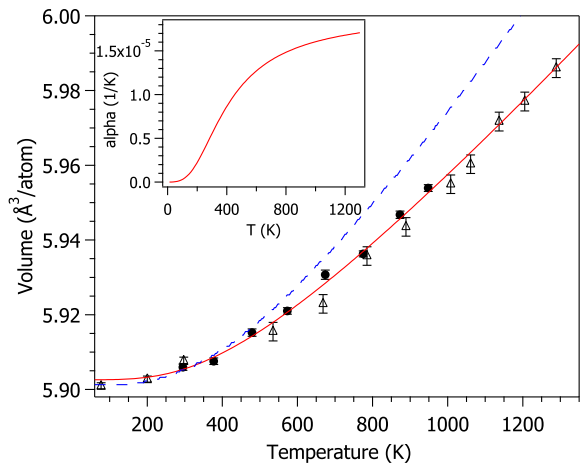}
\caption{\label{RPTE} (Color online) Volume and thermal expansion coefficient as a function of temperature. The solid
circles  are present data and triangles show measurements of Ref.~\onlinecite{Slack75}. The fit to both data sets using
Eq.~(\ref{RPvol}) is shown as the red line. The blue dotted line represents the calculations of Albe\cite{Albe97}. In
the inset, the volume thermal coefficient $\alpha_0$ is represented. }
\end{figure}

We performed volume measurements at room pressure as a function of temperature between 295 and 948 K. The results
are plotted in Fig.~\ref{RPTE} and reproduced in Table~\ref{PVTdata}. They are compared to those of Slack and
Bertram\cite{Slack75} between 77 and 1289 K. The two data sets agree within error bars in the overlapping range,
although we systematically find larger volume values at a given temperature above 300 K. The calculations of
Albe\cite{Albe97} are also shown as the dotted line in Fig.~\ref{RPTE}. They are based on the density functional
theory (DFT) within the local density approximation (LDA) for the static part, and a Debye-Gr\"{u}neisen model
for the thermal part. The calculated volume follows very well the experimental data below 500 K but increasingly
overestimate them above this temperature.

Present and Slack and Bertram\cite{Slack75} data were fitted together using the second-order approximation to the
zero-pressure Gr\"{u}neisen equation of state (see Ref. \onlinecite{Wallacebook} and references therein). In this
approximation, the temperature dependence of the volume is given by:

\begin{equation}\label{RPvol}
V_0(T)=V_{0,0}\left[1+\frac{U_{th}(T)}{Q-bU_{th}(T)}\right]
\end{equation}

where $U_{th}(T)$ is the internal energy due to lattice vibrations, $b=\frac{1}{2}(B'_{0,0}-1)$ and
$Q=(V_{0,0}\,B_{0,0})/\widehat{\gamma}$. $V_{0,0}$, $B_{0,0}$ and $B'_{0,0}$ are respectively the volume, bulk modulus
and its first derivative at zero pressure and temperature. $\widehat{\gamma}$ is the thermobaric Grüneisen parameter
defined by:

\begin{equation}\label{gammatb}
\widehat{\gamma}=-(V/U_{th})(\partial F_{th}/\partial V)_T=(V/U_{th})P_{th}
\end{equation}

where $F_{th}(V,T)$ is the thermal part of the Helmholtz free energy and $P_{th}=-(\partial F_{th}/\partial
V)_T$. If $\widehat{\gamma}$ is independent of $T$, which we assume here, then $\widehat{\gamma}=\gammath{}$
where $\gammath{}$ is the thermodynamic Gr\"{u}neisen parameter ($\gammath{}=(V/C_v)(\partial P/\partial T)_V$
with $C_v$ the specific heat at constant volume)\cite{Wallacebook}. $U_{th}(T)$ was evaluated within the Debye
model:

\begin{equation}\label{DebyeEnergy}
U_{th}(T)=9RT\left(\frac{\Theta_{0}}{T}\right)^3\int_{0}^{\Theta_{0}/T}\frac{x^3}{\exp(x)-1}\,dx
\end{equation}

where $R$ is the ideal gas constant and $\Theta_{0}$ the Debye temperature at $P=0$. For the latter, we used the
value of 1700~K estimated from the infrared spectrum by Gielisse \etal\cite{Gielisse1967}, which is also in
agreement with data on specific heat\cite{Solozhenko1999,Tohei2006}. $B_{0,0}$ was fixed at 397 GPa, which is
obtained by adding to the value of $B_{0}(295\ \mathrm{K})$ found above (Vinet fit), the small correction (2 GPa)
calculated by Albe\cite{Albe97} between 0 and 300~K (Table~\ref{literature}). $B'_{0,0}$ was fixed equal to
$B'_{0}\,(295\ \mathrm{K})=3.62$.

The fit of the $V_0(T)$ data with Eq.~(\ref{RPvol}) then gives  $V_{0,0}=5.9026(4)$~\Vunit and
\gammath{0}=1.04(1) ($\gammath{0}=\gammath{}(P=0)$). It is shown as the solid red line in Fig.~\ref{RPTE}. The
temperature dependence of the thermal expansion coefficient $\alpha_0$, calculated by differentiating
Eq.~(\ref{RPvol}), is plotted in the inset of the same figure. It is seen that Eq.~(\ref{RPvol}) gives a good
reproduction of $V_0(T)$ in the considered temperature range. We also note that the present value for \gammath{0}
is close to the one obtained by Brillouin-zone integration of the phonon dispersion curve ($<\gamma_\nu>=0.95$)
determined by Kern \etal\cite{Kern1999} using density functional theory.

\section{High Pressure and Temperature Equation of state}

\begin{figure}[tb]
\includegraphics[width=8.5cm]{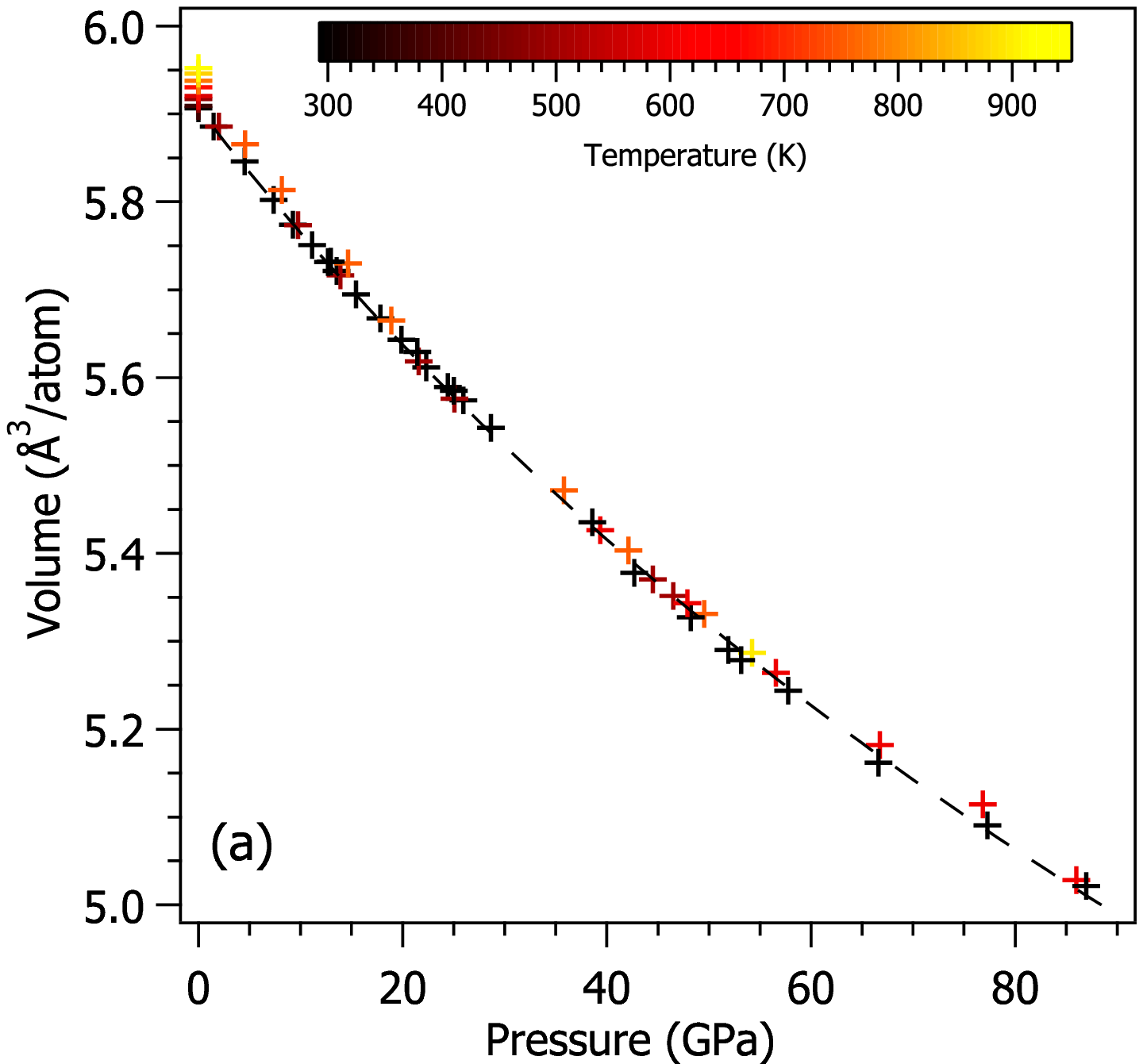}
\includegraphics[width=8.5cm]{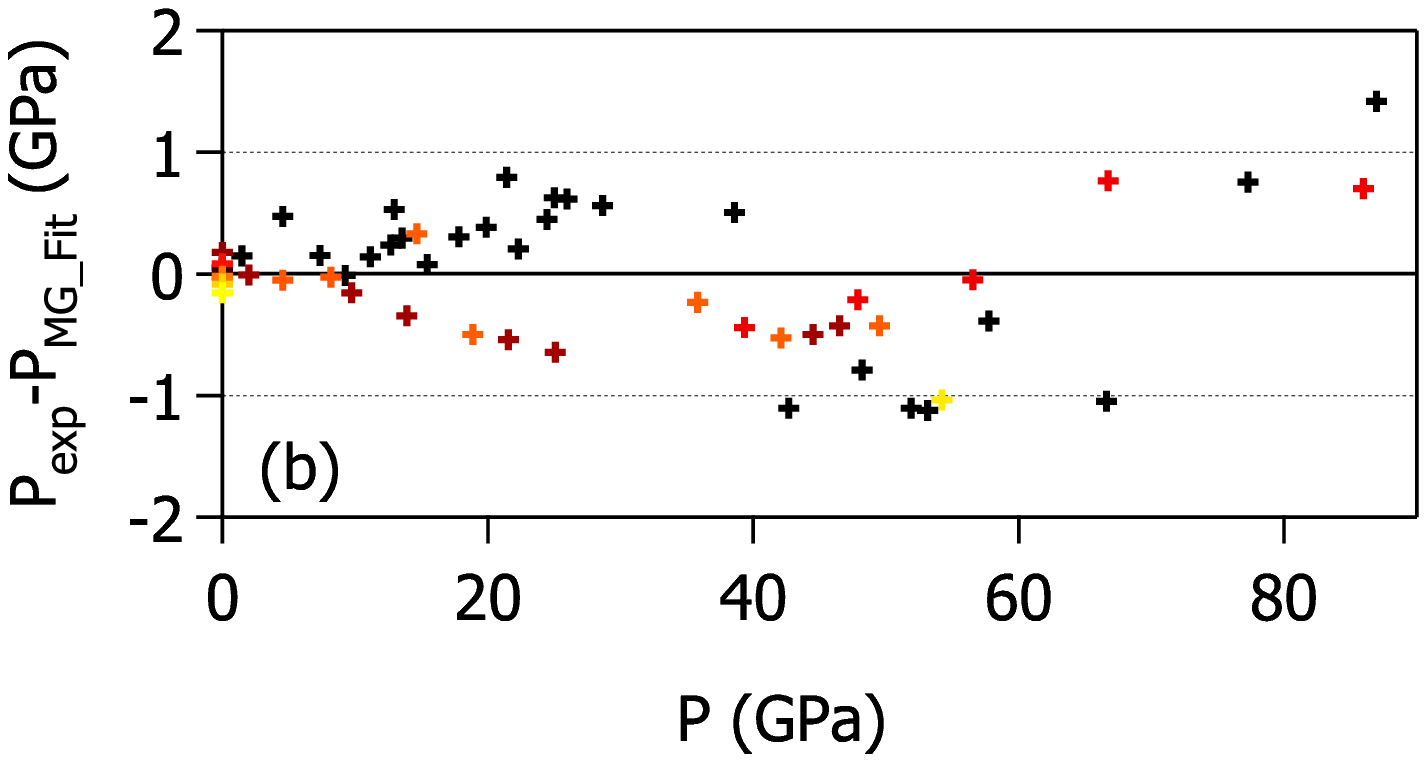}
\caption{\label{HPHTEOS} (Color online) (a) Experimental volume of \cBN as a function of pressure and temperature
obtained in the present work. Crosses show experimental data and the dashed-line is the Vinet fit to the data at
295~K. (b) Difference between experimental pressures and predicted ones using the Mie-Gr\"{u}neisen model with
the parameters given in Table~\ref{MGfit}.}
\end{figure}

High pressure and temperature measurements were performed up to 80 GPa at 500, 600, 750 and 900~K, with the
results listed in Table~\ref{PVTdata}. These are to our knowledge the first reported data at simultaneous HP-HT
conditions. The $P$--$V$--$T$ data set is represented in Fig.~\ref{HPHTEOS}(a).

This data set was least-squares fitted using a Mie-Gr\"{u}neisen model. In this model, the pressure is expressed as the
sum of a "cold" ($P_0$) and thermal ($P_{th}$) part, i.e.:
\begin{equation}\label{MGeq}
P(V,T)=P_0(V) + P_{th}(V,T)
\end{equation}

with $P_{th}(V,T)=0$ at 0~K. To represent $P_0(V)$, a Vinet equation\cite{Vinet1987} is used, whereas the thermal
part is written in the Debye-Gr\"{u}neisen (quasi-harmonic) approximation, using Eqs.
(\ref{gammatb}-\ref{DebyeEnergy}). In this approximation, we have the identity
$\gammath{}=\widehat{\gamma}=\gamma_D$, where $\gamma_D=-\partial \ln(\Theta)/\partial \ln(V)$ is the
Debye-Gr\"{u}neisen parameter. The variation of the Debye temperature $\Theta$ with volume is thus given by that
of $\gammath{}$. Here $\gammath{}$ was allowed to vary with the compression ratio according to the empirical
relation $\gammath{}=\gammath{0}(V/V_{0,0})^q$, where $q$ was taken as constant\cite{Anderson1979}. In the fit,
the only varying parameter is $q$. All the others ($V_{0,0}$, $B_{0,0}$, $B'_{0,0}$, $\gammath{0}$ and
$\Theta_{0}$) were fixed to the values determined as described above. The obtained value for $q$ is $4\pm1.5$ and
the whole parameter set is given in Table~\ref{MGfit}. Fig.~\ref{HPHTEOS}(b) shows the difference between the
experimental pressures and the one predicted by the present model. This difference is for most data points
smaller than $\pm1$ GPa and the rms deviation is 0.6 GPa. If we allow $B_{0,0}$ and $\gammath{0}$ to vary in the
fitting procedure, the resulting values are identical to the starting ones within their standard deviations. The
parameter set given in Table~\ref{MGfit} can thus be regarded as the best one for the Mie-Grüneisen EOS based on
the present data.

\begin{figure}[tb]
\includegraphics[width=8.5cm]{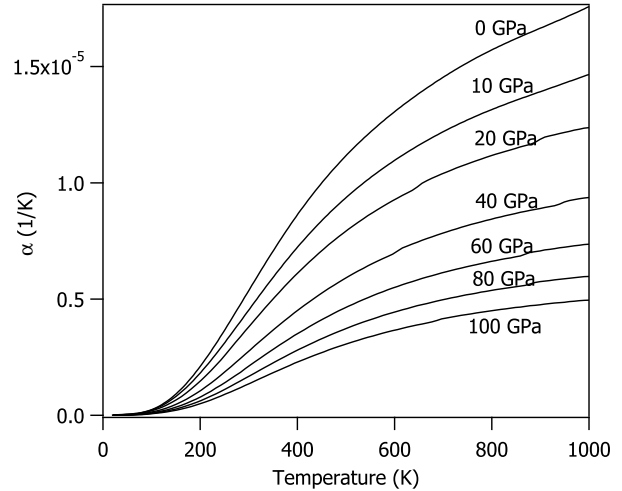}
\caption{\label{alphaVsT} Variation of the thermal expansion coefficient with temperature along several isobars,
as deduced from the Mie-Grüneisen EOS.}
\end{figure}

The value found for $q$ is large compared to the typical range for this parameter
($0.8<q<2.2$)\cite{Anderson1979}. This is to be related to the rapid decrease of the thermal expansion
coefficient with pressure that is observed from present data. Indeed, $\gammath{}$ and $\alpha$ are related by
the thermodynamic relation $\alpha=(1/B)\gammath{}C_V$. In the quasi-harmonic approximation, $\gammath{}$ may be
obtained by the Brillouin-zone integration of the phonon mode Grüneisen parameters. A large reduction of $\alpha$
may thus indicate that the Gr\"{u}neisen parameters for the acoustic modes, which give the dominant contribution,
rapidly decrease with pressure. As a matter of fact, the calculated $\gamma$ for the transverse acoustic branches
at $P=0$ are very low ($\sim 0.25$) along some directions of the Brillouin zone, especially near the zone
boundaries\cite{Kern1999}. In diamond, which presents similar phonon dispersion curves, calculations show a
decrease of these parameters with pressure, eventually leading to negative values of $\gamma$ and $\alpha$ at
ultra-high pressures ($P>700$~GPa)\cite{Xie1999}. We also note that a similar phenomenum is responsible for the
negative thermal expansion in Si at ambient pressure and low temperatures\cite{Xu1991b}. To illustrate the
decrease of $\alpha$ with pressure, we plot in Fig. \ref{alphaVsT} its variation with temperature along several
isobars from 0 to 100 GPa as deduced from the present Mie-Grüneisen EOS.

\begin{table}
\caption{\label{MGfit} Parameters for the \PVT equation of state of \cBN based on a Mie-Gr\"{u}neisen model, with a
Vinet equation\cite{Vinet1987} for the static part and a Debye-Gr\"{u}neisen model for the thermal part
[Eq.~(\ref{MGeq})].}
\begin{ruledtabular}
\begin{tabular}{cccccc}
$V_{0,0}$,\Vunit & $B_{0,0}$, GPa & $B'_{0,0}$ & $\Theta_{0}$, K & \gammath{0} & $q$\\
5.9026(4)        & 397(2)         & 3.62(5)    & 1700          & 1.04(2)     & 4(1.5)\\
\end{tabular}
\end{ruledtabular}
\end{table}

\section{The constrained Raman pressure scale}

\subsection{Temperature dependence of the TO mode Gr\"{u}neisen parameter}

 The present construction of a \PVT EOS enables us to reanalyze our previous
measurements\cite{Datchi2004} of the Raman TO mode at HP-HT. In particular, we are now able to precisely
determine the isothermal mode Gr\"{u}neisen parameter $\gammaTO(T)=-(\partial \ln \nuTO/\partial \ln V)_T$, where
$\nuTO$ is the frequency of the TO mode. To do so, the sample volume was calculated for each $P$--$T$ conditions
at which $\nuTO$ was measured (see \OLC{Datchi2004}) by inverting Eq.~(\ref{MGeq}).

\begin{figure}[tb]
\includegraphics[width=8.5cm]{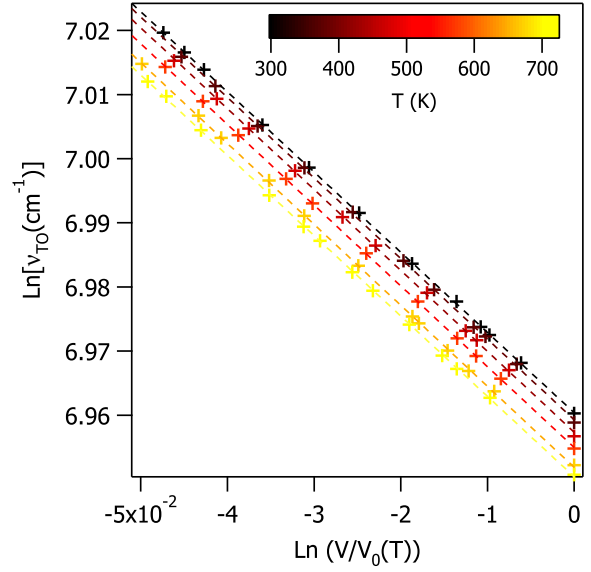}
\caption{\label{LnNuvsLnV} (Color online) Frequencies of the TO Raman mode as a function of $V/V_0(T)$ on a logarithmic
scale. The crosses show experimental data\cite{Datchi2004} and the dashed lines are linear fits.}
\end{figure}

The results are reported in Fig.~\ref{LnNuvsLnV}, where we plot the experimental values of $\ln (\nuTO)$ as a
function of $\ln [V/V_0(T)]$. In the scanned pressure range ($P<21$~GPa), and for each of the five studied
isotherms ($300<T<723$~K), these two quantities appear to be linearly related. This tells us that $\gammaTO(T)$
is constant along each isotherm and may then be directly determined by a linear regression of the data, using the
expression:
\begin{equation}\label{eqgammaTO}
\ln \nuTO(V,T)=-\gammaTO(T) \ln \frac{V(P,T)}{V_0(T)} +\ln \nuTO_0(T)
\end{equation}

where $\nuTO_0(T)=\nuTO(P=0,T)$. The fits to the latter equation are shown as dotted lines in
Fig.~\ref{LnNuvsLnV}. The values of $\gammaTO(T)$ and $\nuTO_0(T)$ so obtained are listed in Table~\ref{gammaTO}.
It can be seen that, within error bars, $\gammaTO(T)$ is constant in this $P$--$T$ range, with an average value
of $1.257(5)$. This value is slightly larger than that given by Aleksandrov \etal\cite{Aleksandrov89} [1.188] and
in good agreement with the theoretical determination of Kern \etal\cite{Kern1999} [1.2].

It can also be observed that $\nuTO_0(T)$ follows $V_0(T)$ in a linear way in the probed $T$ range. Extrapolating
this line to 0~K gives a value of $\nuTO_0$ at zero temperature of 1055.1(1) or 1054.6(2) \invcm , depending on
whether the fitted or experimental $\nuTO_0(T)$ is used.

\begin{table}
\caption{\label{gammaTO} Isothermal mode Gr\"{u}neisen parameter of the Raman TO mode based on the Raman data
reported in \OLC{Datchi2004} and present \PVT data. $V_0$ is the volume at $P=0$ obtained by inverting
Eq.~(\ref{MGeq}). The value of $\nuTO_0\equiv\nuTO(P=0,T)$ predicted by the linear regression of the data
[Eq.~\ref{eqgammaTO}] is also given and compared to the direct measurements. $T$ is in K, $V_0$ in \Vunit and
$\nuTO_0$ in \invcm. The values at $T=0$ are obtained by linear regression of $\nuTO_0(T)$ vs. $V_0(T)$.}
\begin{ruledtabular}
\begin{tabular}{cccccc}
$T$     & $V_0$     &  $\nuTO_0$ (fit)  & $\nuTO_0$ (exp.) & $\gammaTO$ (fit)\\
\hline
300     & 5.9055    &  1054.0(1)          & 1053.93(10)      & 1.251(3)     \\
373     & 5.9084    &  1052.9(2)          & 1052.43(20)      & 1.256(5)       \\
473     & 5.9139    &  1050.8(2)          & 1050.17(13)      & 1.262(6)      \\
573     & 5.9209    &  1048.3(1)          & 1048.17(27)      & 1.259(2)      \\
673     & 5.9290    &  1045.1(1)          & 1045.45(13)      & 1.262(4)      \\
723     & 5.9334    &  1043.6(1)          & 1043.91(41)      & 1.253(4)      \\
\hline
0       & 5.9026    &  1055.1(1)          & 1054.6(2)           &            \\
\end{tabular}
\end{ruledtabular}
\end{table}

\subsection{The high-temperature pressure scale}

In Ref.~\onlinecite{Datchi2004} we recalled the various reasons that make \cBN a good candidate for pressure
measurement in a DAC at high temperature. A pressure scale was given based on the measurements of the TO mode
frequency ($\nuTO$) at high $P$--$T$. The form of this pressure scale derives from the first-order Murnaghan EOS,
and is obtained by inverting the following equation for the P-T dependance of $\nuTO$:

\begin{equation}
\nuTO(P,T)=\nuTO_0(T)\left(1+\frac{B'_0}{B_0(T)}P\right)^{\gamma_{TO}/B'_0} \label{Murnaghan}
\end{equation}

In Ref.~\onlinecite{Datchi2004}, $B_0(T)$ and $B'_0$ were considered as fit parameters, taking for $\gammaTO$ the
value reported by Ref.~\onlinecite{Aleksandrov89}. We assumed then that $\gammaTO$ was independent of
temperature, which is confirmed by the present work. Our EOS allows us to better constrain Eq.~(\ref{Murnaghan})
by imposing physical constraints on the various parameters. Since this pressure scale is to be used for high
temperatures, we only consider variations above 300 K. The variation of $B_0$ with temperature was deduced from
our Mie-Grüneisen EOS by fitting isotherms at 100~K intervals by a Vinet equation. A quadratic form is found
suitable to represent the behavior of $B_0$ between 300 and 2000 K, with the following expression:
\begin{eqnarray}\label{B0vsT}
B_0(T>300\,\mathrm{K})&=396.5(5)-0.0288(14)\,(T-300)\\\nonumber & -6.84(77)\ten{-6}\,(T-300)^2
\end{eqnarray}

The data for $\nuTO(P,T)$ was then fitted to Eq.~(\ref{Murnaghan}), using Eq.~(\ref{B0vsT}) as the expression for
$B_0(T)$. Adding a temperature dependence to $B'_0$ does not improve the fit, so we kept it as constant. The
value of $\gammaTO=1.257$ obtained above was used. As in Ref.~\onlinecite{Datchi2004}, $\nuTO_0(T)$ was expressed
as a quadratic form, where the coefficients are allowed to vary within the boundaries given by
Ref.~\onlinecite{Herchen93}. We thus obtain a revised pressure scale, reading as:

\begin{equation}\label{Pscale}
P=(B_0(T)/3.62)\left\{\left[\frac{\nuTO(P,T)}{\nuTO_0(T)}\right]^{2.876}-1\right\}
\end{equation}

with $B_0(T)$ as in Eq.~(\ref{B0vsT}) and $\nuTO_0(T)=1058.4(5)-0.0091(23)\,T-1.54(22)\times\ten{-5}\,T^2$

\begin{figure}[tb]
\includegraphics[width=8.5cm]{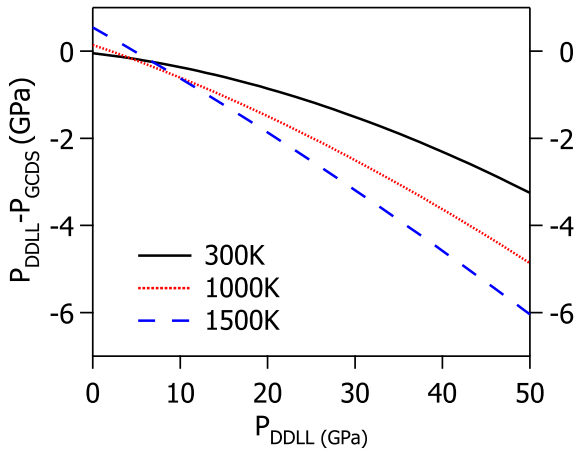}
\caption{\label{CompPscale} (Color online) Comparison between the present pressure scale [Eq.~(\ref{Pscale})] and the
one given by Goncharov \etal [Ref.~\onlinecite{Goncharov2005}, Eq.~(2)].}
\end{figure}

Recently, Goncharov \etal\cite{Goncharov2005} reported Raman measurements up to 1750 K and 40 GPa in argon
pressure medium and proposed a pressure scale in a  form similar to Eq.~(\ref{Pscale}). The two scales are
compared in Fig.~\ref{CompPscale} in the pressure range 0-50 GPa at 300, 1000 and 1500 K. The difference
increases with pressure and temperature and reaches $\sim$10\% at 50 GPa and 1000 K. The fact that the present
scale is constrained by measured physical parameters gives us confidence that it can be safely used in the 0-100
GPa and 300-1000 K range. \PVT as well as Raman data to higher $T$ would be desirable to extend the calibration.

\section{conclusions}

We have reported experimental \PVT data on \cBN from x-ray diffraction experiments in a resistively-heated
diamond anvil cell. Volume measurements extend to 160 GPa at 295~K and 80~GPa at 500-900~K. To our knowledge,
these are the first reported EOS data at simultaneous HP-HT conditions. By fitting the room-temperature EOS to
various EOS models, we extracted the values of the bulk modulus [395(2) GPa] and its first-pressure derivative
[3.62(5)]. We have also examined the variation of these parameters with respect to the chosen calibration of the
ruby standard, and showed that the one recently proposed by Holzapfel\cite{Holzapfel2005} (H2005) provides very
good consistency between static compression data and \emph{ab initio} predictions. Coupling the present
information with independent measurements of $B'_0$, such as obtained by sound-propagating experiments, would be
of great interest to better constrain the ruby standard calibration in the megabar range. A good description of
our full data set was obtained using a Mie-Grüneisen model, where the thermal pressure originating from lattice
vibrational energy is calculated in the Debye approximation. We observe a rapid decrease of the thermal expansion
coefficient with pressure, which itself reflects on the strong variation of the thermal Grüneisen parameter. The
Mie-Grüneisen EOS was then used to determine the mode Grüneisen parameter of the Raman TO mode, which was found
to be temperature independent in the range 300-723~K. A new formulation of the Raman pressure scale was then
deduced from the present results, which should hold in the $P$-$T$ range 0-100~GPa and 300-1000~K. Extension of
this work to higher temperatures would be valuable to better constrain the thermal effects, and extend the
calibration of the Raman scale.

\acknowledgments

We thank V. L. Solozhenko for providing us with \cBN samples,  B. Canny (IMPMC) for his help in preparing the
experiments, and M. Mezouar (ESRF ID27) for his help during the x-ray diffraction runs. The authors acknowledge
the European Synchrotron Radiation Facility for provision of synchrotron radiation facility during beamtime
allocated to proposal HS-2514.

\bibliographystyle{apsrev}

\end{document}